\def\aa{A\&A}               %Astronomy & Astrophysics%
\def\anj{AJ}                %Astronomical Journal%
\documentclass{aa}
\usepackage{natbib}

\begin{document}

\title{Adaptive optics imaging of low and intermediate redshift\\ quasars
\thanks{Based on data obtained at the Canada-France-Hawaii telescope
which is operated by CNRS of France, NRC of Canada and the University of
Hawaii.}}
%\subtitle{}

\author{Isabel M\'arquez\inst{1} \and Patrick Petitjean\inst{2,3} 
  \and Bertrand Th\'eodore\inst{4,5} \and Malcolm~Bremer\inst{6} 
  \and Guy Monnet\inst{7} \and Jean-Luc Beuzit\inst{8} 
}

\offprints{I. M\'arquez \email{isabel@iaa.csic.es}}

\institute{Instituto de Astrof\'{\i}sica de Andaluc\'{\i}a (CSIC),
Apdo. 3004, 18080 Granada (Spain)
\and 
Institut d'Astrophysique de Paris, CNRS, 98bis Bd Arago, 
F-75014 Paris, France 
\and
UA CNRS 173 -- DAEC, Observatoire de Paris-Meudon, F-92195 Meudon
Cedex, France 
\and
Service d'A\'eronomie du CNRS, BP 3, F-91371 Verri\`ere le Buisson, 
France
\and
ACRI, 260 route du Pin Montard, BP 234, F-06904 Sophia-Antipolis,
France 
\and
Bristol Univ. (Dept. of Physics)
H H Wills Physics Laboratory, Tyndall Av, Bristol BS8 1TL, United Kingdom
\and
European Southern Observatory, Karl Schwarschild Stra$\ss$e 2,
D-85748 Garching-bei-M\"unchen, Germany
\and
Canada-France-Hawaii Telescope Corporation, 65-1238 Mamaloha Highway, Kamuela, HI 96743, USA
} 
\date{Received / Accepted}

\authorrunning{M\'arquez et al.}
\titlerunning{Adaptive optics imaging of low and intermediate redshift QSOs}

\abstract{
We present the results of adaptive-optics imaging in the H and K bands
of 12 low and intermediate redshift ($z$~$<$~0.6) quasars using the
PUEO system mounted on the Canada-France-Hawaii telescope. Five
quasars are radio-quiet and seven are radio-loud. The images,
obtained under poor seeing conditions, and with the QSOs ($m_{\rm
V}$~$>$~15.0) themselves as reference for the correction, have typical
spatial resolution of $FWHM$~$\sim$~0.3~arcsec before deconvolution. The
deconvolved H-band image of PG~1700$+$514 has a spatial resolution
of 0.16~arcsec and reveals a wealth of details on the companion and
the host-galaxy.\\
Four out of the twelve quasars have close companions and obvious signs
of interactions. The two-dimensional images of three of the host-galaxies unambiguously reveal bars and spiral arms. The morphology of
the other objects are difficult to determine from one dimensional surface brightness profile and deeper images are needed.\\
Analysis of mocked data shows that elliptical galaxies are always
recognized as such, whereas disk hosts can be missed for small disk scale 
lengths and large QSO contributions.
}

\maketitle

\section{Introduction}

Evidence that nearby bright galaxies contain massive dark objects in
their center has become increasingly compelling over the last few
years and early suggestions that a tight correlation exists between
the mass of the dark object and the mass of the bulge \citep{kr95}
have been convincingly corroborated \citep{magorrian,ferrarese}.  It
is thus possible that AGN activity is a usual episode of the history
of most, if not all, present-day bright galaxies.  One way to
investigate this is to determine the luminosity and morphology of
galaxies hosting quasars. In addition, this gives clues on the range
of conditions needed for strong nuclear activity to occur.
%and
%It is therefore of prime importance to
%detect and measure the luminosity of the host-galaxies of AGNs and
%quasars. This should be done at any redshift as the evolution of their
%masses should tell us how the most massive objects in the Univese
%formed (Haehnelt \& Kauffman 2000).
\par\noindent 
Recent observations of host-galaxies with HST have questioned the
previous belief that radio-loud and radio-quiet quasars are found
preferentially in the center of, respectively, elliptical and spiral
galaxies \citep{bahcall97,mclure99}. This supports previous
conclusions from IR ground-based observations
\citep{dunlop,mcleod95a,taylor} that the distinction in host-galaxy
characteristics between the two classes of QSOs is subtle \citep[see also]
[for larger redshift]{hooper}. More surprising is the
finding by McLure et al. (1999) that host-galaxies of luminous AGNs
(both radio-quiet and radio-loud) are all massive ellipticals whereas
Bahcall et al. (1997) found that, in the same redshift range,
host-galaxies are of various morphologies but that all radio-loud
quasars have bright elliptical hosts or occur in interacting systems
\citep{kirkhakos}.  \par\noindent The detection and analysis of
host-galaxies is difficult even from space
\citep{bahcall94,bahcall95,mcleod95b, mcleod00}.  Indeed, the
determination of the PSF and the subtraction of the point source image
are crucial in this work. Differenciation between the two classical
profiles, either an exponential disk or a de Vaucouleur power-law, is
effective only in the regions close to the center, or in the far-wings
of the PSF \citep[see Fig.~1 of][]{mclure00}. This is the main
limiting factor in determining host-galaxy morphologies from the
ground.  With the advent of adaptive optics, it will be possible to
alleviate the seeing limitation
\citep{stockton98,aretxaga,hutchings99}. Observing in the infrared
minimizes the difference in luminosity between the host and nucleus
again improving our ability to determine the host morphology.

PSF determination is still a major problem but the difficulties are
balanced by the prospect of using very soon 10~m class telescopes
which will provide higher sensitivity and better spatial resolution.
\par\noindent In this paper we present the result of a pilot
programme aiming at testing the capabilities of adaptive optics in
this field. We present the data in Section~2, discuss each object in
Section~3, analyze the results in Section~4 and conclude in Section~5.

\section{Sample selection and data}\label{sample}

\begin{center}
\begin{table*}[h!]\caption{Log of the observations.}\label{obs}
\begin{tabular}{lcccccc}
\hline
\\
     Name      & Other names & $m_{\rm v}$  &  $z$   &run$^1$ &
Exposure & FWHM$^2$\\
               &             &        &        &        &
(minutes)& (arcsec)\\
\hline
\\
Q~0955+326   & 3C 232  & 15.8 & 0.530 & 1   & 20    & 0.36      \\
PG 1001+291  &         & 16.0 & 0.329 & 1   & 60    & 0.24      \\
PG 1012+008  &         & 15.6 & 0.185 & 1   & 20    & 0.26      \\
PKS 1302-102 &         & 15.2 & 0.286 & 1,2 & 20,48 & 0.26,0.42 \\
PG 1402+261  &         & 15.5 & 0.164 & 1,2 & 40,64 & 0.26,0.34 \\
B2 1425+26   &         & 15.7 & 0.366 & 1,2 & 35,64 & 0.24,0.34 \\
Q~1618+177   & 3C 334  & 16.0 & 0.555 & 1   & 30    & 0.36      \\
PG 1700+518  &         & 15.1 & 0.290 & 1,2 & 40,40 & 0.26,0.30 \\
Q~1704+608   & 3C 351  & 15.3 & 0.371 & 1,2 & 40,64 & 0.26,0.43 \\
B2 1721+34   &         & 15.4 & 0.206 & 1   & 75    & 0.24      \\
PG 2112+059  &         & 15.6 & 0.466 & 1,2 & 30,32 & 0.24,0.34 \\
PKS~2128-12  &         & 16.2 & 0.501 & 2   & 24$^a$& 0.61,0.48 \\
\\
\hline
\end{tabular}
\begin{footnotesize}

$^1$ Run 1: may 98 (H-band), Run 2: may 99 (K-band)\\
$^2$ Measured on the available stars observed close in time to the objects\\
$^a$ Also 28 min. in H 
\\
\end{footnotesize}
\end{table*}
\end{center}
In order to use adaptive optics correction quasars were selected such
that the nuclei were bright enough to be used as the wavefront
reference point source. The sample of radio-quiet quasars were all PG
quasars with m$_b$ $<$ 16.5 and with redshift less than 0.6. The radio-loud
objects were selected from 3C, 4C, B2 and PKS catalogues with the
same magnitude and $z$ criteria. The final objects observed (see Table 1) were
selected based upon the suitability for the observing conditions on
the observing runs. 
%In other words, the objects with the
%brightest nuclei, at the lowest $z$ were given priority.

We used the CFHT adaptive optics bonnette (PUEO) and the 
IR camera KIR on May 1998 (run 1) and May 1999 (run 2). 
The weather conditions were poor during both runs and the FWHM of the seeing 
PSF was never better than 0.8 arcsec. The adaptative-optics correction was
performed on the QSOs themselves. 
The quasar was centered successively in the center of the four
quadrants of the detector. The exposure time for individual images was
two minutes.
The background was determined by
median--averaging the frames and the flat--field was taken to be
the normalized dark-substracted background. 
The images were then aligned and added. The final images have a typical 
resolution of $FWHM$~$\sim$~0.3 arcsec. After each science observation
an image of a star with similar magnitude as the QSO 
was taken in order to determine the PSF and use it to deconvolve the images.
Due to rapid variations in the wheather conditions however, it was not
always possible to follow this predefined procedure. 
\par\noindent
A synthetic PSF function, derived from the stellar images was used to
deconvolve each of the images.  As it was not always possible to apply
a standard procedure due to fluctuating seeing conditions, a careful
although, somewhat arbitrary choice of the PSF had to be done.  In
Fig.~\ref{pg1700} we show the images of PKS~1700+514 obtained using,
for the deconvolution, three different PSFs from stars observed during
the same night. These have respectively, $FWHM$~= 0.30,
0.42 and 0.48 arcseconds.  The initial image of the object has a
resolution of $FWHM$~=~0.26~arcsec and the star observed just after
the science exposure has $FWHM$~=~0.48~arcsec.  It is apparent that
the best result is obtained using the star with the FWHM closest
to that of the science exposure.  Here, we were guided in the exercice
by the existence of the HST image by Hines et al. (1999). In general,
this illustrates the crucial role played by a careful PSF
determination in AO observations.
\par\noindent
Results are summarized in Table~2.  Columns
$\#3$ and $\#4$ give, respectively, the number of objects (probably
companions) found within
5~ and 10~arcsec from the quasar down to $m_{\rm H}$~=~20.5; columns
$\#5$ and $\#6$ give the maximum radial distance (in arcsec and 
kilo-parsec) to which the host is
detected at a significance level of 3$\sigma$ above the background;
column $\#7$ gives the total
magnitude of the object in the H-band
and columns $\#8$ and $\#9$ those of the host-galaxy as derived from
the PSF subtraction and profile fitting respectively (see Section~4);
the assigned morphology, which comes from the 2D brightness
distribution and the comparison of the two profile fittings, is given
in column $\#10$.

\section{Comments on individual objects}\label{objects}

\subsection{PKS~1302--102}\label{pks1302}
\begin{figure}[h]
\vspace{13cm}
%\special{psfile=ms1059f1.gif hoffset=0 voffset=0 hscale=50 vscale=50 angle=0}
%\rule{0.4pt}{10cm}% line thickness, height of picture
\caption{Images of PKS 1302-102 in the H-band. 
The top-left panel and bottom panel correspond to the image,
respectively, before and after deconvolution. The spatial resolutions
are respectively $FWHM$~$\sim$~0.32~ and 0.24~arcsec.  The top right
panel shows the two companions after substraction of the PSF and a
model for the host-galaxy. The inset in the bottom panel corresponds
to a higher contrast version of the inner 4~arcsec$\times$4~arcsec.}
\label{pk1302}
\end{figure}
The image obtained at CFHT under moderately good seeing conditions
is of similar quality to that obtained with HST by Bahcall et al. (1995)  
\citep[see for comparison][]{hutchings94} with $FWHM$~=~0.24~arcsec 
after deconvolution (see Fig.~\ref{pk1302}). 
The two objects at 1 and 2~arcsec from the
quasar are well-detached, and are more clearly seen when both the PSF and
a model for the host-galaxy (obtained by masking the companions and
fitting ellipses to the isophotes) are subtracted. It is unlikely that
these companions are intervening objects as strong associated metal line
absorption would be expected at such small impact parameter when no
such absorption is detected in the HST spectrum down to $w_{\rm
obs}$~$\sim$~0.2~\AA~ \citep{jannuzi}. The host-galaxy of this
quasar has been detected by HST and fitted with a $r^{1/4}$ profile
\citep{disney}.
M\'arquez et al. (1999) derive that the galaxy contributes 40\% the
total flux in the J-band when McLeod \& Rieke (1994) measure this
contribution to be 31\% of the total flux in the H-band after fitting 
an exponential profile to the host-galaxy.  
We have performed a similar fit on the present data and found that
the contribution of the galaxy amounts to 39\% in H and 18\% in K (see
Table \ref{res}). However, Fig. \ref{profils} shows that an 
$r^{1/4}$ profile is a better fit.
In that case, the contribution of the host-galaxy to the
total light is 70\% (60\%) in H (K), in good
agreement with the values derived by subtracting a scaled version of
the PSF and directly integrating the residual flux.

\par\medskip\noindent
\subsection{PG~1700+514}\label{Pg1700}
\begin{figure}[h]
\vspace{13cm}
%\special{psfile=ms1059f2.gif hoffset=0 voffset=0 hscale=50 vscale=50 angle=0}
%\rule{0.4pt}{13cm}% line thickness, height of picture
\caption{Images of PG~1700+514 after deconvolution using
the PSF given by three stars observed during the same night.
The best deconvolution is obtained using the star with the FWHM 
closest to that of the quasar. The resulting image (bottom) has a final 
resolution of $FWHM$~=~0.16~arcsec. 
}
\label{pg1700}
\end{figure}
PG~1700+514 is one of the most infrared-luminous, radio-quiet BAL
quasar \citep{turnshek85,turnshek97}. Ground-based imaging revealed an
extension about 2~arcsec north-east of the quasar \citep{stickel}
which was shown by adaptive-optics imaging and follow-up spectroscopy
to be a companion with a redshift 140~km~s$^{-1}$ blueward of the
quasar \citep{stockton98}. NICMOS observations lead Hines et al. (1999) to
argue that the companion is a collisionally induced ring galaxy.  The
fit to the SED and the Keck spectrum of the companion imply that the
light is emitted by an old population of stars plus a 85 Myr old
star-burst \citep{canalizo}. Note however that the H-band
flux ($m_{\rm H}$~$\sim$~16.6) deduced from HST imaging is much larger
than that predicted by the model.  Stockton et al. (1998) showed
that the inclusion of embedded dust can produce a spectral-energy
distribution that is consistent with both the optical
spectrophotometry and the IR photometry.
\par\noindent
The image obtained at CFHT is shown in Fig.~\ref{pg1700}. 
We confirm the findings by Stockton et al. (1998) that the companion has the
appearance of an arc with several condensations.  We used different
PSF to deconvolve the image. The best deconvolution is obtained using
the star with the FWHM closest to that of the AGN (0.30
arcsec, see Section 2). The image has a final resolution of 0.16 arcsec and is
probably the best image obtained yet on this object.  The companion is
seen as a highly disturbed system with a bright nucleus and a
ring-like structure; the nucleus beeing decentered with respect to the
ring. The host-galaxy is clearly seen around the quasar with a bright
extension to the south-west, first noted by Stickel et al. 1995 and 
clearly visible in the optical images by Stockton et al. (1998). In addition, 
we detect a bright knot to
the south-east which is not seen in the NICMOS data probably because
of the presence of residuals in the PSF subtraction.  
The comparison between the HST and CFHT images of PG~1700+514
shows how powerful AO can be, and bodes well for the use of the technique 
on 10~m-class telescopes.
No obvious relation is found between the near-IR image and the 
radio map \citep{hutchings92}.

\begin{center}
\begin{table*}[h!]\caption{Characteristics of host-galaxies}\label{res}
\begin{tabular}{lcccccllll}
\hline
\\
     Name      &  $z$   & \multicolumn{2}{c}{Companions} & 
\multicolumn{2}{c}{Extension} &  m$_T^H$(m$_T^K$) & m$_{PSFsub}^H$(m$_{PSFsub}^K$) & m$_{r^{1/4}}^H$(m$_{r^{1/4}}^K$) & Host type\\
               &        & 5 arcsec & 10 arcsec & arcsec & kpc$^a$  &  & \\
\hline
\\
Q~0955+326     & 0.530 &   &   & 2.2 & 7.9 & 14.18         & 14.25         & 12.56         & E \\
PG 1001+291    & 0.329 &   &   & 2.9 & 8.1 & 13.87         & 14.10         & 14.84         & SBa \\
PG 1012+008    & 0.185 & 1 & 1 & 3.0 & 6.3 & 16.71         & 16.97         & 17.19         & E-Sa \\
PKS 1302-102   & 0.286 & 2 &   & 2.6 & 7.1 & 13.45 (12.53) & 13.84 (12.89) & 13.84 (13.08) & E \\
PG 1402+261    & 0.164 &   &   & 3.0 & 5.3 & 13.23 (11.91) & 14.07 (12.86) & 14.07 (11.67) & SBa \\
B2 1425+26     & 0.366 &   & 2 & 2.3 & 7.5 & 14.26 (13.30) & 14.69 (14.01) & 15.01 (14.23) & E-Sa \\
Q~1618+177     & 0.555 & 1 & 2 & 1.5 & 5.4 & 14.68         & 15.18         & 15.48         &  \\
PG 1700+518    & 0.290 & 1 &   & 2.5 & 6.7 & 12.90 (11.75) & 13.89 (12.50) & 13.89 (12.15) & Sa? \\
Q~1704+608     & 0.371 &   & 1 & 2.3 & 6.9 & 13.30 (12.21) & 14.37 (12.89) & 14.30 (12.84) & E \\
B2 1721+34     & 0.206 &   & 2 & 1.5 & 3.2 & 13.95         & 14.13         & 15.07         &  \\
PG 2112+059    & 0.466 & 1 & 1 & 2.0 & 6.6 & 13.64 (12.64) & 14.10 (12.96) & 14.71 (13.51) & E? \\
PKS~2128-12    & 0.501 &   & 1 & 1.7 & 6.1 & 13.42 (12.77) & 14.39 (13.33) & 14.50 (13.21) &  \\
\\
\hline
%
%\begin{footnotesize}
%
%\medskip
\multicolumn{8}{l}{$^a$ $H_{\rm o}$~=~100~km~s$^{-1}$~Mpc$^{-1}$; 
$q_{\rm o}$~=~0.5}\\
%\end{footnotesize}
\end{tabular}
\end{table*}
\end{center}
\subsection{Other objects}\label{others}
\begin{figure*}[h]
\vspace{22cm}
%\special{psfile=ms1059f3.gif hoffset=0 voffset=0 hscale=80 vscale=80 angle=0}
%\rule{0.4pt}{20cm}% line thickness, height of picture
\caption{Images of objects described in Section~3.3 and Tables~1 and 2. 
They are K band images for PKS1302-102 and PKS2128-12 and H band images for 
the rest. The scale is marked with the horizontal bar in the top-left figure and is the same for all of them.}
\label{images}
\end{figure*}
\par\medskip\noindent
{\bf 3C~232 -- TON~0469} This object has attracted much interest
because it is located 2~arcmin ($\sim$~8$h^{-1}$~kpc at $z$~=~0.0049)
north of the nearby ($z$~=~0.0049) galaxy NGC~3067 and has been
considered as the prototype of the galaxy--quasar physical association
as a probe for anomalous redshifts \citep{hoyle}. The HST
spectrum shows a strong Mg~{\sc ii} system at $z_{\rm abs}$~=~0.0049
\citep{tumlinson} which has been shown to be due to an H~{\sc i}
tidal tail originating in the disk of the foreground galaxy
\citep{carilli}. The MERLIN map at 1.6~GHz shows a slight extension to
the east \citep{akujor}. The host-galaxy is detected in the CFHT
image with an extension to the south-west up to 5~arcsec from the
quasar). From both the image (Fig. \ref{images}) and the profile
fitting (Fig. \ref{profils}) we derive that the host is better
described as an elliptical galaxy.
\par\medskip\noindent
{\bf PG 1001+291 -- Ton~0028} Only one image from the HST archive has
been published yet \citep{boyce99}. The latter authors favor the
interpretation that the object is an interacting object with two
nuclei one 1.9~arcsec to the south-west and the other 2.3~arcsec to
the north-east.  We find no evidence for the presence of two nuclei. In
particular there is no residual after subtraction of a 
host-galaxy model. The profile is not well fitted by a single disk
component, due to the presence of a bump which is characteristic of
the bar component in the profiles of barred spirals. We therefore
conclude that the elongated NE--SW feature seen in our image probably
reveals the presence of a strong bar in an otherwise spiral galaxy.
\par\medskip\noindent
{\bf PG 1012+008} This radio-quiet quasar interacts with two
companions located to the north and east \citep{heckman}. It has
been imaged with HST by Bahcall et al. (1997) who concluded from
2D-modelling that the host is a spiral galaxy with
$r_{1/2}$~=~10.8~kpc. They also fit the one-dimensional profile as
an elliptical with $r_{1/2}$~=~24.5~kpc. The best 2D fit 
by McLure et al. (1999)
is a large elliptical with $r_{1/2}$~=~23~kpc. We
find that the best fit to the H-band profile is a $r^{1/4}$ law with a
scalelength of 2.9 arcsec or 6$h^{-1}$ kpc (see Fig.~\ref{images}).
\par\medskip\noindent
{\bf PG 1402+261 -- TON~0182} Bahcall et al. (1997) determined from HST
imaging that the host of this quasar is a bright spiral galaxy with
prominent H~{\sc ii} regions located along the spiral arms. The CFHT
image shows that there is a bright elongation extending
$\sim$~2~arcsec away from the nucleus on both sides of it in the
NW--SE direction, probably tracing a strong bar also seen
as a bump in the surface brightness profile, see Fig. \ref{profils}.
\par\medskip\noindent
{\bf B2 1425+26 -- TON~0202} This quasar is hosted by a bright galaxy
probably in mild interaction with one of the three companions located
within 10~arcsec from the point source and listed by
Hutchings et al. (1984) and Block \& Stockton (1991). Two of these
companions are detected in our H and K-band images. The host-galaxy
has been classified as elliptical by previous investigators
\citep{malkan,kirkhakos}. However, the isophotes in our
image are distorted and an arc-like feature is clearly seen to the
west \citep[see also the F555W image of][]{kirkhakos}. Moreover, an
emission line nebulosity is detected by H$\alpha$, H$\beta$, [O~{\sc
iii}] emission up to 3~arcsec from the nucleus
\citep{boroson,stockton87}. Our fit shows that an elliptical galaxy
(or an early spiral) is a good description for the host-galaxy (see
Fig. \ref{profils}).
\par\medskip\noindent
{\bf 3C~334} The host-galaxy has twisted isophotes. As noted by
Lehnert et al. (1999), the galaxy appears elongated approximatively in
the same direction as the radio structure \citep{hutchings98}. Note
that the arc-like structure seen to the south in the deconvolved image
has been detected by its [OII] emission \citep{hes}.  Comparison
of the new AO corrected images with previous ground-based images
\citep[e.g.][]{hes,marquez} clearly shows the power
of AO technique.  We note that three Lyman-$\alpha$ absorptions are
detected in the quasar spectrum with $w_{\rm obs}$~=~0.37, 1.00 and
0.21~\AA~ at redshifts $z_{\rm abs}$~=~0.5387, 0.5449 and 0.5491,
slightly smaller than the emission redshift $z_{\rm em}$~=~0.555
\citep{jannuzi}. Spectroscopy of the companions is required to
determine if they are somehow associated with these absorption
systems. The profile fitting does not allow to distinguish between a
disk-like or an elliptical host.
\par\medskip\noindent
{\bf 3C~351} We observed this field in the H and K-bands (see
Fig.~\ref{images}).  The two objects closest to the line of sight in
our image are situated 7~arcsec north-east of the quasar. One is a
bright spiral galaxy, well resolved in our images, the other is very
faint (it is barely visible in Fig. \ref{images} but it is more
clearly seen in the deconvolved image) and could be a companion of the
former. Lanzetta et al. (1995) and Le~Brun et al. (1996) have searched
the field around 3C~351 for galaxies responsible for Lyman-$\alpha$
and C~{\sc iv} absorption observed by Bahcall et al. (1993) in the HST
spectrum of the quasar. There are two Lyman-$\alpha$ absorption lines
at $z_{\rm abs}$~=~0.2216 and 0.2229, the former showing C~{\sc iv}
absorption as well. Le~Brun et al. (1996) identified the metal line
system with a galaxy 710~$h^{-1}_{50}$~kpc away from the line of
sight. The redshift of the two objects closest to the quasar have not been
determined however and it is possible that these two objects are responsible
for the absorptions. If they are at $z$~=~0.222, the impact parameter
is of the order of only 15~$h^{-1}$~kpc. It is very important to
confirm this for our understanding of the nature of H~{\sc i} halos
around low-$z$ galaxies.  Boyce et al. (1998) detect a companion at
about 3.3~arcsec east to the quasar.  We do not detect
this companion in any of our images although it is within the
possibilities of our imaging. This may indicate that the flux in the
HST-F702W filter is dominated by line emission or that the object is very
blue.  As there is no strong intervening absorption in the spectrum of
the quasar, it is probable that this companion is at the redshift of
the quasar. Note that 3C~351 exhibits a strong associated system with
H~{\sc i} Lyman-$\alpha$, C~{\sc iv}, N~{\sc v} and O~{\sc vi}
absorptions.  The presence of the associated system does not seem to
be related to any other imaging property of the QSO.
\par\medskip\noindent
{\bf B2 1721+34} The host-galaxy is detected up to 1.5~arcsec from the
central point-source in the H-band CFHT image (see
Fig.~\ref{images}). The one-dimensional profile shows that a $r^{1/4}$
law fits the galaxy profile better than an exponential fit, but images
with better S/N ratio are needed to confirm this at higher
significance.
\par\medskip\noindent
{\bf PG 2112+059} Two companions are detected within 10~arcsec from
the nucleus.  An elliptical profile
better describes the light distribution of the host-galaxy.
\par\medskip\noindent
{\bf PKS~2128-12} Disney et al. (1995) fitted the one dimensional
profile with a de Vaucouleur law of radius $R_{\rm e}$~=~37.4~kpc.
The images of this object have the poorest resolution in our sample.
We are therefore unable to determine a reliable fit to the host-galaxy
profile.  We detect a companion at 7 arcsec north-east to the
nucleus.

\section{Analysis}\label{analysis}
\begin{figure*}[h]
\vspace{22cm}
\includegraphics{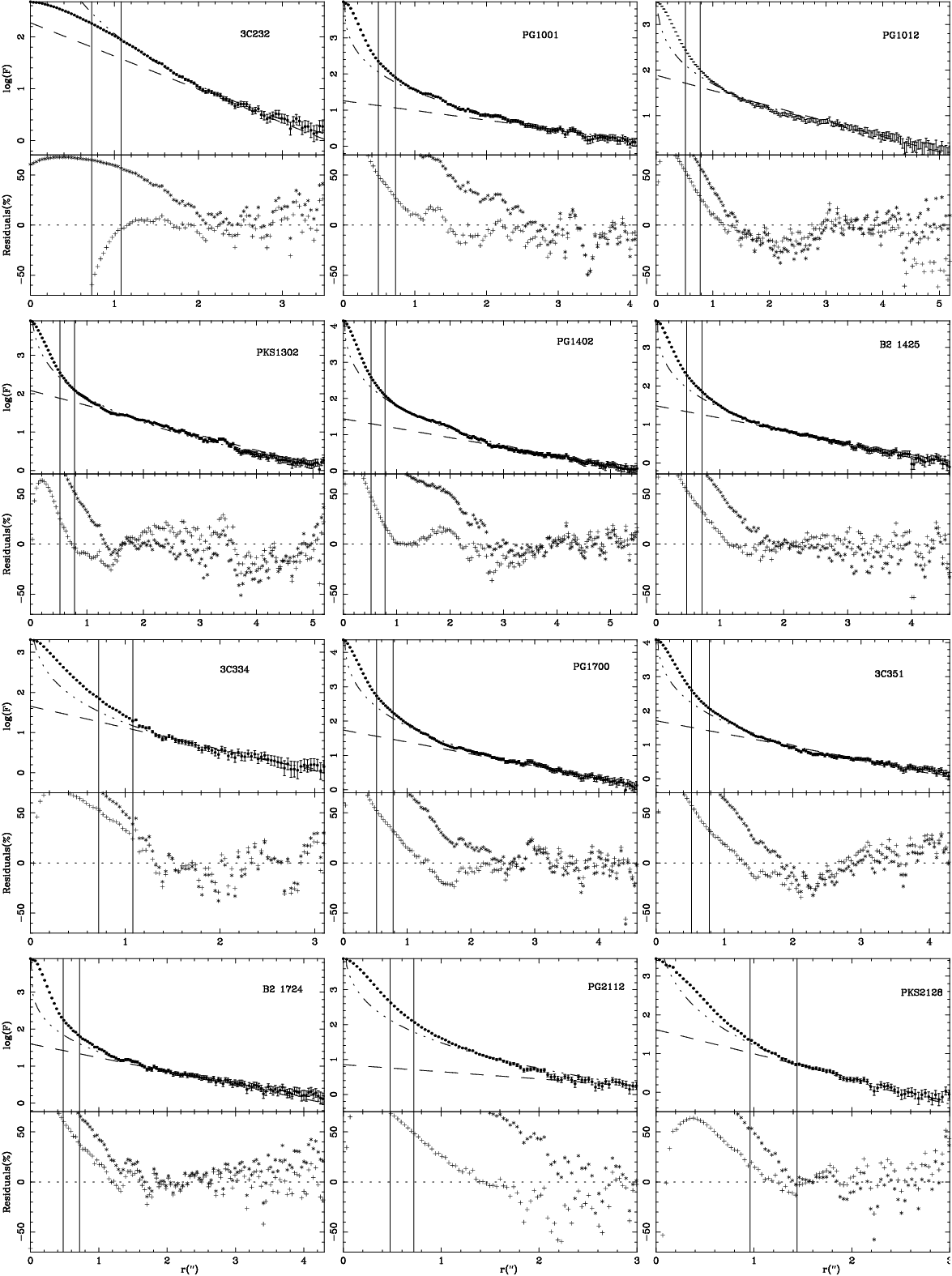}
%\rule{0.4pt}{22cm}% line thickness, height of picture
\caption{{\sl Top}: Surface brightness profiles (dots and
error bars) obtained with the
undeconvolved images in the H-band (except for PKS~2128-12, for
which we used the K-band image because of its better PSF).
Fits by a disk (dashed line) and a r$^{1/4}$ law
(dotted-dashed line) are overplotted. {\sl Bottom}: Residuals from the 
disk (stars) and r$^{1/4}$ (crosses) fits in percentage. 
The vertical lines are drawn at respectively two and three times the PSF 
FWHM.}
\label{profils}
\end{figure*}

In each of the images, we first masked out the companion objects and
the ghosts due to the telescope.  We then obtained the surface
brightness profiles of the galaxies using the IRAF\footnote{IRAF is 
the Image Analysis and Reduction Facility made available to the 
astronomical community by the National Optical Astronomy
Observatories, which are operated by the Association of Universities
for Research in Astronomy (AURA), Inc., under contract with the
U.S. National Science Foundation.} task {\sl ellipse}.  The resulting
profiles were fitted over the radius range from 3 times the FWHM of
the PSF up to the point where the galaxy surface brightness level
falls below 2 $\times \sigma$ of the background level.  We have
systematically fitted an exponential disk and a de Vaucouleurs
r$^{1/4}$ law.  The results are shown in Fig.~\ref{profils}, where the
residuals from the subtraction of the fitted profiles to the real
profiles are also plotted.  Out of the 10 objects for which we can
extract some morphological information, two of the host-galaxies are
most probably barred spirals, the rest being ellipticals or very early
type spirals (see Table 2).  We note that in general the number of
points we can use to fit either profile is not large enough to
unambiguously distinguish between the two fitted profiles. At small
radii, the excess of light between the observed profile and the model
disk can be due to the presence of a bulge. The morphology indicated
in column \#10 of Table~2 is determined considering both the 2D
luminosity spatial distribution and the 1D profile. It is apparent
that to discriminate between both morphologies, the S/N ratio must be
high at large radii.
\par\noindent
\par\noindent
The magnitudes of the hosts have been derived by integrating the
$r^{1/4}$ profiles for all the objects. Indeed, it can be seen
on Fig.~\ref{profils} that there is an excess of light at small radius
compared to the disk profile for all objects. This suggests that
in our sample, the disk galaxies have also a strong bulge and/or
a strong bar. This is confirmed by the 2D luminosity distribution.
Results are given in Table \ref{res}. We have 
also subtracted a scaled version of the most suitable PSF for each
nucleus \citep[imposing a non-negative profile in the center, see][]{marquez}
The resulting host magnitudes, computed by
integrating the PSF-subtracted images, are in good agreement
with those obtained from the profile fitting (see Table~2).
\par\noindent
In order to test our fitting procedure, we have generated images of
model elliptical and disk galaxies with scale-lengths and effective
surface brightness within the range derived from the data.  The same
orientation and axis ratio is given to all of them. A point source
is added in the center of the galaxy to mimic the quasar.
An appropriate amount of noise is added, and then the images are 
convolved with a typical observed PSF. The mocked images are analyzed
 in the same way as real data.
\par\noindent
\begin{figure*}[h]
\vspace{6cm}
\includegraphics{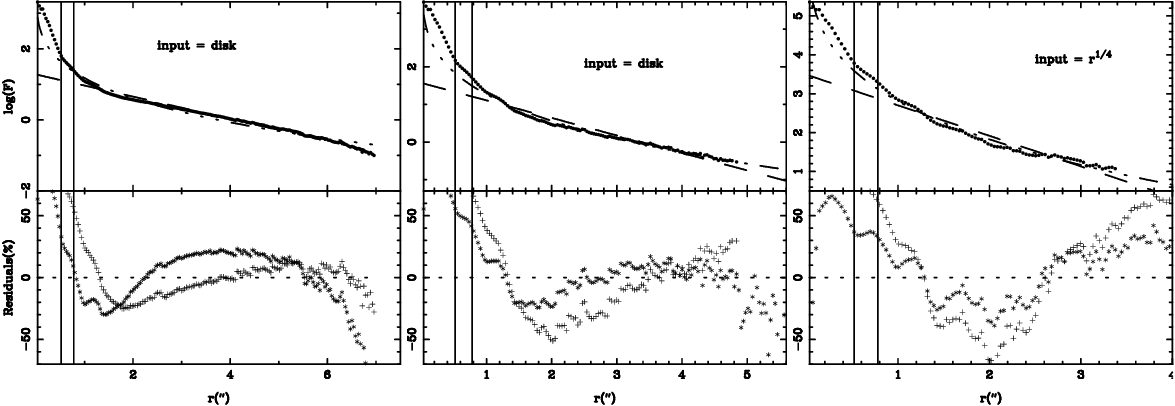}
%\rule{0.4pt}{6cm}% line thickness, height of picture
\caption{Illustrative examples of mocked images and their fits. The top panel 
is the surface brightness profiles with models overplotted
(exponential disk as a solid line and $r^{1/4}$ law as a dashed line).
(a) The input is a disk-galaxy and the mocked image is better fitted
by a disk host (the quasar contributes 25\% to the total light).  (b)
The input is a disk-galaxy but the mocked image is equally well fitted
by the two laws. The disk parameters are the same as is (a), but the
quasar contributes 70\% to the total light. (c) When the input is an
elliptical galaxy, the mocked image is always better fitted by a
$r^{1/4}$ profile.}
\label{prof_modeles}
\end{figure*}
\par\noindent
\begin{figure}[h]
\vspace{6cm}
\includegraphics{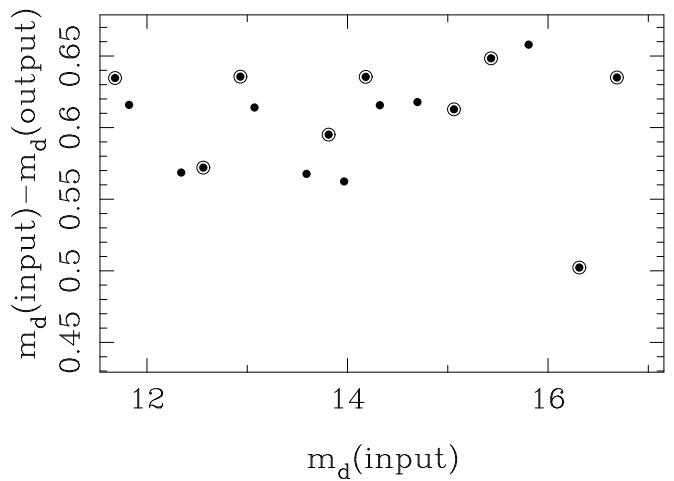}
%\rule{0.4pt}{6cm}% line thickness, height of picture
\caption{Test of the fitting procedure on disk-galaxies. We plot here
the difference between the input and output magnitudes versus the
input magnitude in the case
the host-galaxy has the same magnitude as the quasar.
A  dot is surrounded by a circle means that  
the fit by an r$^{1/4}$ law is at least as good as the fit by
an exponential disk.
}
\label{mag_disque}
\end{figure}
\par\noindent
\begin{figure}[h]
\vspace{6.5cm}
\includegraphics{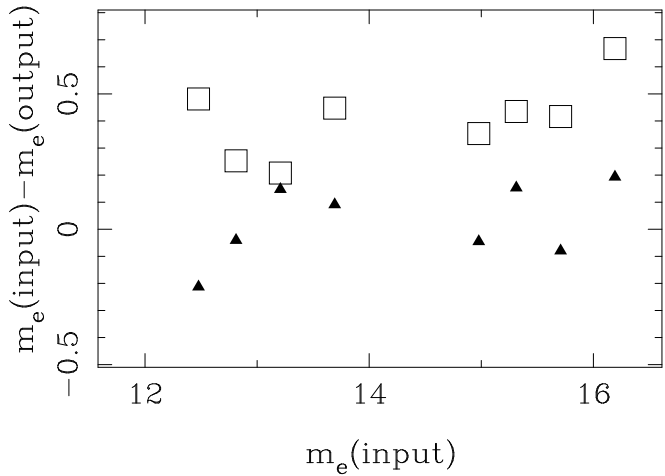}
%\rule{0.4pt}{6cm}% line thickness, height of picture
\caption{Test of the fitting procedure on elliptical galaxies. We plot here
the difference between the input and output magnitudes versus the
input magnitude. 
Triangles correspond to the systems where the QSO contributes 
to the total luminosity as the host-galaxy does, squares represent those 
cases in which the host luminosity is half that of the QSO.}
\label{mag_ellip}
\end{figure}
\par\noindent
We first note that an elliptical galaxy is always recognized as
an elliptical galaxy by the fitting procedure, whereas a disk-galaxy
is better fitted by a $r^{1/4}$ law when the unresolved 
point-source contributes more than half the total light.
This is illustrated in Fig.~\ref{prof_modeles}. This means that,
at least with data of similar quality to those presented here, 
the fraction of elliptical galaxies in the sample may be overpredicted. 
Going deeper, at least 0.5 to 1 magnitude, should help solve this problem 
as it is apparent that the distinction between spiral and 
elliptical profiles is easier when the galaxy is detected at larger distances
from the central point-source. 
\par\noindent
It is interesting to note that the output magnitudes are brighter than
the input in both cases, elliptical or disk galaxies (see
Figs.~\ref{mag_disque},\ref{mag_ellip}). The reason for this is
probably the difficulty in determining the extension of the PSF wings
which, if not subtracted properly, will artificially increase the flux
of the host-galaxy. In the case of spirals, the difference is as large
as 0.6 magnitudes when the contribution of the point-source is the
same as the contribution of the host-galaxy (see
Figs.~\ref{mag_disque}).  For the ellipticals, the difference is less
but still important when the QSO dominates the total flux.

\par\noindent

Note that the ratio between the QSO and the host-galaxy luminosities 
is expected to increase with redshift. The above bias tends to imply
that host-galaxy luminosities could be overestimated. 
%This goes toward reinforcing the result by Kukula et al. (2000)
%that host-galaxies are fainter at higher redshift.
		 
\section{Conclusions}\label{conclu}

Adaptive-optics imaging in the H and K bands has been used to study
the morphology of QSO host-galaxies at low and intermediate redshifts
($z$~$<$~0.6).  We detect the host-galaxies in 11 out of 12 quasars,
of which 5 are radio-quiet and 7 are radio-loud quasars. The images,
obtained under poor seeing conditions, and with the QSOs themselves as
reference for the correction, have typical spatial resolution of
$FWHM$~$\sim$~0.3~arcsec before deconvolution. In the best case, the
deconvolved H-band image of PG~1700$+$514 (with a spatial resolution
of 0.16~arcsec) reveals a wealth of detail on the companion and the
host-galaxy, and is probably the best-quality image of this
 object thus far.

Four of the quasars in our sample have close companions and show
obvious signs of interactions. The two-dimensional images of three of
the host-galaxies unambiguously reveal bars and spiral arms.
For the other objects, it is difficult to determine the host-galaxy 
morphology on the basis of one dimensional surface brightness fits alone.

We have simulated mocked images of host-galaxies, both spirals and
ellipticals, and applied the same analysis as to the data.  Disk hosts
can be missed for small disk scalelengths and large QSO
contributions. In this case, the host-galaxy can be misidentified as
an elliptical galaxy. Elliptical galaxies are always recognized as
such, but with a luminosity which can be overestimated by up to 0.5
magnitudes.  The reason for this is that the method used here tends to
attribute some of the QSO light to the host. This is also the case for
disk galaxies with a strong contribution of the unresolved component.

\begin{acknowledgements}
We acknowledge the suggestions by the referee, Alan Stockton, that 
helped improving the presentation.
I.~M\'arquez acknowledges financial support from the Spanish
Ministerio de Educaci\'on y Ciencia (EX94-8826734). This work is
financed by DGICyT grants PB93-0139 and PB96-0921. 
Financial support to
develop the present investigation has been obtained through the Junta
de Andaluc\'{\i}a TIC-114. 

\end{acknowledgements}

%%%%%%%%%%%%%%%%%%%%%%%%%%%%%%%%%%%%%
\end{document}